# High Sensitivity Terahertz Detection through Large-Area Plasmonic Nano-Antenna Arrays

**Nezih Tolga Yardimci[1], Mona Jarrahi[1]**

[1]Electrical Engineering Department, University of California Los Angeles
Los Angeles, CA 90095, United States

*Correspondence to: mjarrahi@ucla.edu

**ABSTRACT**
Plasmonic photoconductive antennas have great promise for increasing responsivity and detection sensitivity of conventional photoconductive detectors in time-domain terahertz imaging and spectroscopy systems. However, operation bandwidth of previously demonstrated plasmonic photoconductive antennas has been limited by bandwidth constraints of their antennas and photoconductor parasitics. Here, we present a powerful technique for realizing broadband terahertz detectors through large-area plasmonic photoconductive nano-antenna arrays. A key novelty that makes the presented terahertz detector superior to the state-of-the art is a specific large-area device geometry that offers a strong interaction between the incident terahertz beam and optical pump at the nanoscale, while maintaining a broad operation bandwidth. The large device active area allows robust operation against optical and terahertz beam misalignments. We demonstrate broadband terahertz detection with signal-to-noise ratio levels as high as 107 dB.

**INTRODUCTION**
Time-domain terahertz spectroscopy systems offer unique functionalities for various imaging and sensing applications [1-8]. Electro-optic sampling is a widely used technique for detecting terahertz pulses in time-domain terahertz spectroscopy systems [9-13]. It measures changes in polarization state of an optical beam interacting with terahertz radiation in an electro-optic crystal. Despite its capability to offer broad detection bandwidths, electro-optic sampling requires complex optical setups and tight beam alignment, which have hindered its usage in portable terahertz imaging and sensing systems [14, 15]. An alternative approach for detecting terahertz pulses is using an optically pumped photoconductive antenna, comprised of an ultrafast photoconductor integrated with a terahertz antenna [16-20]. When a terahertz beam is incident on a photoconductive antenna, a terahertz electric field is induced across the photoconductor contact electrodes, drifting photo-generated carriers and inducing a photocurrent proportional to the incident terahertz field. Terahertz detection with photoconductive antennas does not require any additional optical component, offering a very promising platform for portable terahertz imaging and sensing systems. However, responsivity and detection sensitivity of conventional photoconductive antennas is bound by low drift velocity of the photo-generated carriers in semiconductor substrates, limiting the induced photocurrent in response to an incident terahertz beam.

It has been recently shown that utilizing plasmonic contact electrodes is very effective in enhancing the responsivity and detection sensitivity of photoconductive antennas [21-24]. The use of plasmonic contact electrodes increases the optical pump intensity and photo-generated carrier concentration in their close proximity. This reduces the average transport path length of the photo-generated carriers to the contact electrodes. Therefore, significantly higher photocurrent levels are induced in response to an incident terahertz beam. However, the detection bandwidth of previously demonstrated plasmonic photoconductive antennas has been limited by bandwidth constraints of their terahertz antennas and photoconductor parasitics [21-24]. In order to maintain broad operation bandwidths, previously demonstrated plasmonic photoconductive antennas have been realized with very small active areas to minimize capacitive parasitics of the device. This imposes very tight constraints on optical focusing and alignment to achieve acceptable detection sensitivities. This work presents large-area plasmonic photoconductive nano-antenna arrays as a powerful solution for addressing the limitations of previously demonstrated plasmonic photoconductive antennas used in terahertz time-domain spectroscopy systems.

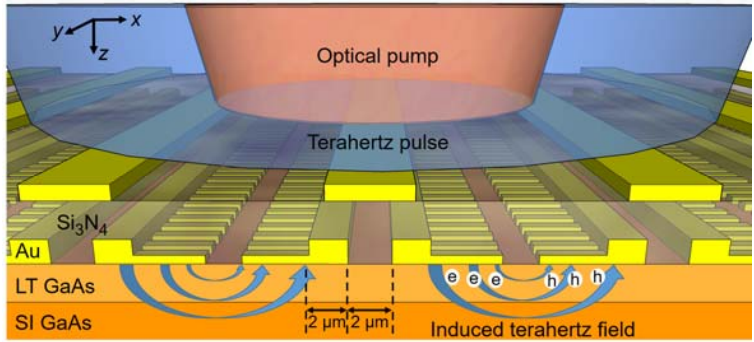

**Figure 1: Terahertz detection through plasmonic photoconductive nano-antenna arrays.** Schematic diagram and operation principles of plasmonic photoconductive nano-antenna arrays used for detecting pulsed terahertz radiation.

Schematic diagram and operation principles of the presented terahertz detector are illustrated in Fig. 1. The device consists of plasmonic photoconductive nano-antenna arrays fabricated on a short carrier lifetime substrate. Dipole-type nano-antennas and a low-temperature grown (LT) GaAs substrate are used for the specific device implementation discussed in this paper. The overall size of the nano-antenna arrays can be chosen arbitrarily large to facilitate efficient coupling of the optical pump beam and terahertz radiation. The nano-antennas are connected together in the vertical direction (along *y* axis) and separated from each other in the horizontal direction (along *x* axis), as illustrated in Fig. 1. Their geometry is chosen to enhance the induced terahertz field between the nano-antenna tips when illuminated with an *x*-polarized terahertz beam. In the meantime, the length of the nano-antennas is chosen to be much smaller than terahertz wavelengths to achieve broadband operation. The nano-antenna geometry is also chosen to allow excitation of surface plasmon waves when illuminated with a *y*-polarized optical pump beam [25, 26]. This concentrates the majority of the photo-generated carriers in close proximity to the nano-antennas [21, 22]. Additionally, a $Si_3N_4$ antireflection coating layer is used to maximize optical transmission into the LT-GaAs substrate. An array of Au stripes on top of the $Si_3N_4$ layer is used to shadow the horizontal gaps between the nano-antennas. This prevents inducing a photocurrent in the opposite direction to that of the plasmonic photoconductive nano-antenna arrays [27]. By reducing the distance of the majority of the photo-generated carriers from the nano-antennas and enhancing the induced terahertz fields that drift the photo-generated carriers, significantly higher responsivity and detection sensitivity levels can be achieved.

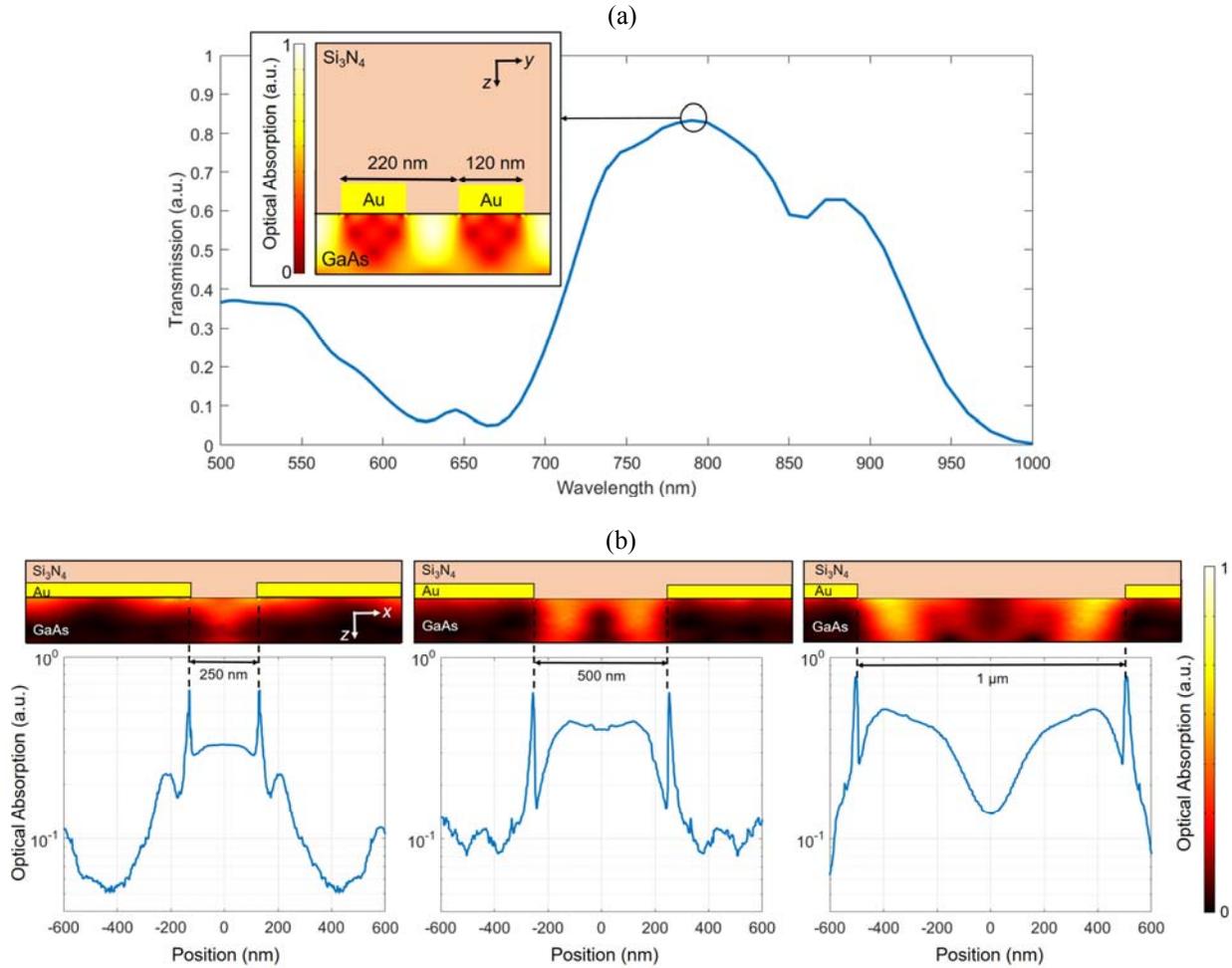

**Figure 2: Optical characteristics of the plasmonic nano-antennas.** (a) Optical transmission spectrum through the plasmonic nano-antennas, inset shows optical absorption inside the substrate at a vertical cross-section of the nano-antennas at 800 nm pump wavelength. (b) Optical absorption between the nano-antenna tips at a 10 nm depth inside the substrate, the inset color plots show optical absorption profile within a 150 nm depth inside the substrate.

A Finite Element Method-based multi-physics software package (COMSOL) is used to analyze the interaction of an *x*-polarized terahertz and *y*-polarized optical pump beam with the designed plasmonic nano-antennas. The analysis shows that Au gratings with 220 nm periodicity, 120 nm width, and 50 nm height, covered by a 290 nm thick $Si_3N_4$ antireflection coating layer offer 84% optical transmission into the LT-GaAs substrate at 800 nm wavelength (Fig. 2a). Since optical transmission into the substrate is accompanied by excitation of surface plasmon waves, optical absorption is significantly enhanced at the edges of the plasmonic nano-antennas (Fig. 2a inset). This enhancement factor is independent of the nano-antenna tip-to-tip gap size since the surface plasmon waves are excited by a *y*-polarized optical pump beam (Fig. 2b). However, the nano-antenna tip-to-tip gap size directly impacts the optical absorption between the nano-antenna tips.

While optical absorption between the nano-antenna tips is diffraction limited for a nano-antenna tip-to-tip gap size of 250 nm, it is significantly increased when the tip-to-tip gap size is increased to 500 nm and 1 μm (Fig. 2b inset).

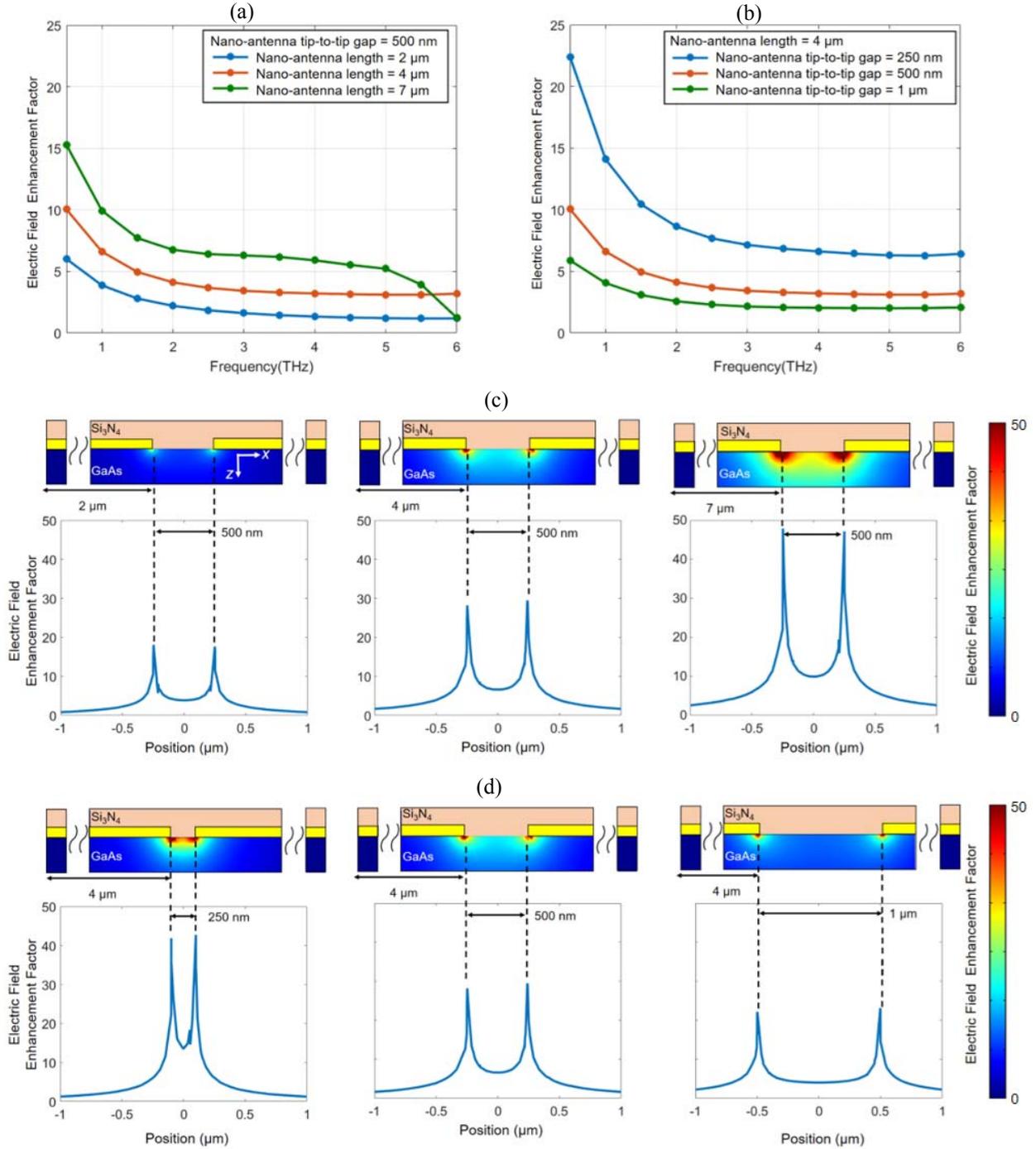

**Figure 3: Terahertz characteristics of the plasmonic nano-antennas.** (a) Electric field enhancement factor in the middle of the nano-antenna tips as a function of the nano-antenna length and tip-to-tip gap size is shown in (a) and (b), respectively. Electric field enhancement factor between the nano-antenna tips at 1 THz as a function of the nano-antenna length and tip-to-tip gap size is shown in (c) and (d), respectively.

The length of the nano-antennas has a direct impact on the device operation bandwidth as well as terahertz field enhancement inside the device active area. Since the vertical gap size between the adjacent nano-antennas (220 nm) is much smaller than terahertz wavelengths, an *x*-polarized terahertz beam cannot transmit through the nano-antenna arms. It can efficiently transmit through the tip-to-tip gaps between the nano-antennas if $\lambda^{THz}_{eff} > \sqrt{\varepsilon_{r-sub}}\Lambda_x$, where $\lambda^{THz}_{eff}$ represents the effective wavelength of the incident terahertz beam inside the substrate, $\varepsilon_{r-sub}$ represents the relative permittivity of the substrate, and $\Lambda_x$ represents the nano-antenna periodicity along the *x*-axis [28]. Under these circumstances, the induced terahertz field between the nano-antenna tips is proportional to $\Lambda_x/g$, where *g* represents the nano-antenna tip-to-tip gap size [28]. Therefore, higher terahertz field enhancement factors are offered by nano-antennas with smaller tip-to-tip gap sizes and larger lengths. The induced terahertz field is significantly reduced if $\lambda^{THz}_{eff} < \sqrt{\varepsilon_{r-sub}}\Lambda_x$. As a result, the device operation bandwidth is limited when using long nano-antennas [28].

Figure 3a shows the terahertz field enhancement factor in the middle of the nano-antenna tips as a function of frequency for different nano-antenna lengths. It illustrates the tradeoff between the field enhancement factor and operation bandwidth as a function of the nano-antenna length. While the field enhancement factor increases as the nano-antenna length increases from 2 μm to 7 μm, it starts to roll off above 5 THz for a nano-antenna length of 7 μm. Figure 3b shows the terahertz field enhancement factor in the middle of the nano-antenna tips as a function of frequency for different nano-antenna tip-to-tip gap sizes. As expected, the field enhancement factor is inversely proportional to the nano-antenna tip-to-tip gap size. Since terahertz coupling is through excitation of surface waves along the nano-antenna arms, higher field enhancement factors are achieved in close proximity to the nano-antenna tips (Figs. 3c and 3d).

In order to achieve high-sensitivity terahertz detection, the nano-antenna geometry should be selected such that the induced terahertz field and optical pump intensity are maximized in close proximity to the nano-antenna tips. By maximizing the concentration of the photo-generated carriers and the terahertz electric field drifting the photo-generated carriers, higher photocurrent levels are induced and higher detection sensitivity levels are achieved, consequently. Therefore, designs with longer nano-antennas are preferred as far as the required detection bandwidths are supported by the device ($\lambda^{THz}_{eff} > \sqrt{\varepsilon_{r-sub}}\Lambda_x$). However, the optimum nano-antenna tip-to-tip gap size depends on different factors. On one hand, narrower tip-to-tip gaps offer stronger terahertz field enhancement factors inside the device active area. On the other hand, wider tip-to-tip gaps result in higher photocarrier concentration levels between the nano-antenna tips, where the highest terahertz field intensities are induced.

**RESULTS**
The impact of the nano-antenna geometry on the device performance is also explored experimentally. Terahertz detector prototypes with 0.5×0.5 mm² area, nano-antenna lengths of 2 μm and 4 μm, and nano-antenna tip-to-tip gap sizes of 250 nm, 500 nm, and 1 μm are fabricated on a LT-GaAs substrate with 0.3 ps carrier lifetime (Fig. 4).

A time-domain terahertz spectroscopy setup is used to characterize performance of the fabricated detector prototypes under the same operation conditions. A large-area plasmonic photoconductive terahertz source is used to generate terahertz pulses in the time-domain terahertz spectroscopy setup [27]. 135 fs wide pulses from a Ti:sapphire mode-locked laser with 76 MHz repetition rate and 800 nm wavelength are used to pump the terahertz source and detector prototypes. The optical beam spot size is adjusted to 300 μm and its polarization is set to be normal to the nano-antennas. The polarization of the incident terahertz pulses on the detector prototypes is set to be parallel to the nano-antennas.

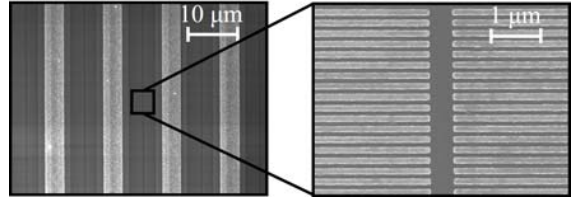

**Figure 4: Fabricated plasmonic nano-antenna array.** Scanning electron microscope image of a plasmonic nano-antenna array fabricated on a LT-GaAs substrate with a nano-antenna length of 2 μm and tip-to-tip gap size of 500 nm.

Figure 5a shows the measured photocurrent by a detector prototype with a nano-antenna length of 4 μm and tip-to-tip gap size of 500 nm in the time-domain. For each time-domain photocurrent data point, 35 measurements are taken and averaged with an integration time of 10 ms. Photocurrent pulses with 180 nA peak level and 0.45 ps FWHM width are measured at 200 mW optical pump power. Figure 5b shows the detected power spectra obtained from the time-domain photocurrent data. A drop in the detector response is observed at optical pump powers exceeding 250 mW. This is due to the carrier screening effect caused by high photocarrier concentration levels at high optical pump powers.

Performance of devices with shorter nano-antennas is less affected by the carrier screening effect. This is because the number of the photo-generated carriers and the induced terahertz fields inside the device active area are smaller for devices with shorter nano-antennas. As illustrated in Fig. 5c, while the responsivity of the detector prototype with a nano-antenna length of 4 μm and tip-to-tip gap size of 500 nm rolls off at optical pump powers above 250 mW, the responsivity of a similar detector prototype with a nano-antenna length of 2 μm does not show any drop up to 450 mW optical pump power. Despite the negative impact of the carrier screening effect, higher responsivity levels are offered by the detector with a nano-antenna length of 4 μm at optical pump powers below 350 mW. This is due to the larger number of photocarriers and higher terahertz field levels induced inside the device active area.

As discussed in previous sections, narrower nano-antenna tip-to-tip gaps offer stronger terahertz field enhancement factors but result in lower photocarrier concentration levels inside the device active area. This tradeoff can be clearly observed when comparing the responsivity of detector prototypes with the nano-antenna length of 2 μm and tip-to-tip gap sizes of 250 nm, 500 nm, and 1 μm (Fig. 5d). Although the highest terahertz field enhancement factors and photocarrier concentration levels are offered by the nano-antennas with tip-to-tip gap sizes of 250 nm and 1 μm, respectively, the highest responsivity levels are achieved by the nano-antenna with a tip-to-tip gap size of 500 nm. Moreover, the device with a nano-antenna tip-to-tip gap size of 250 nm is less affected by the carrier screening effect due to low photocarrier concentration levels inside the device active area.

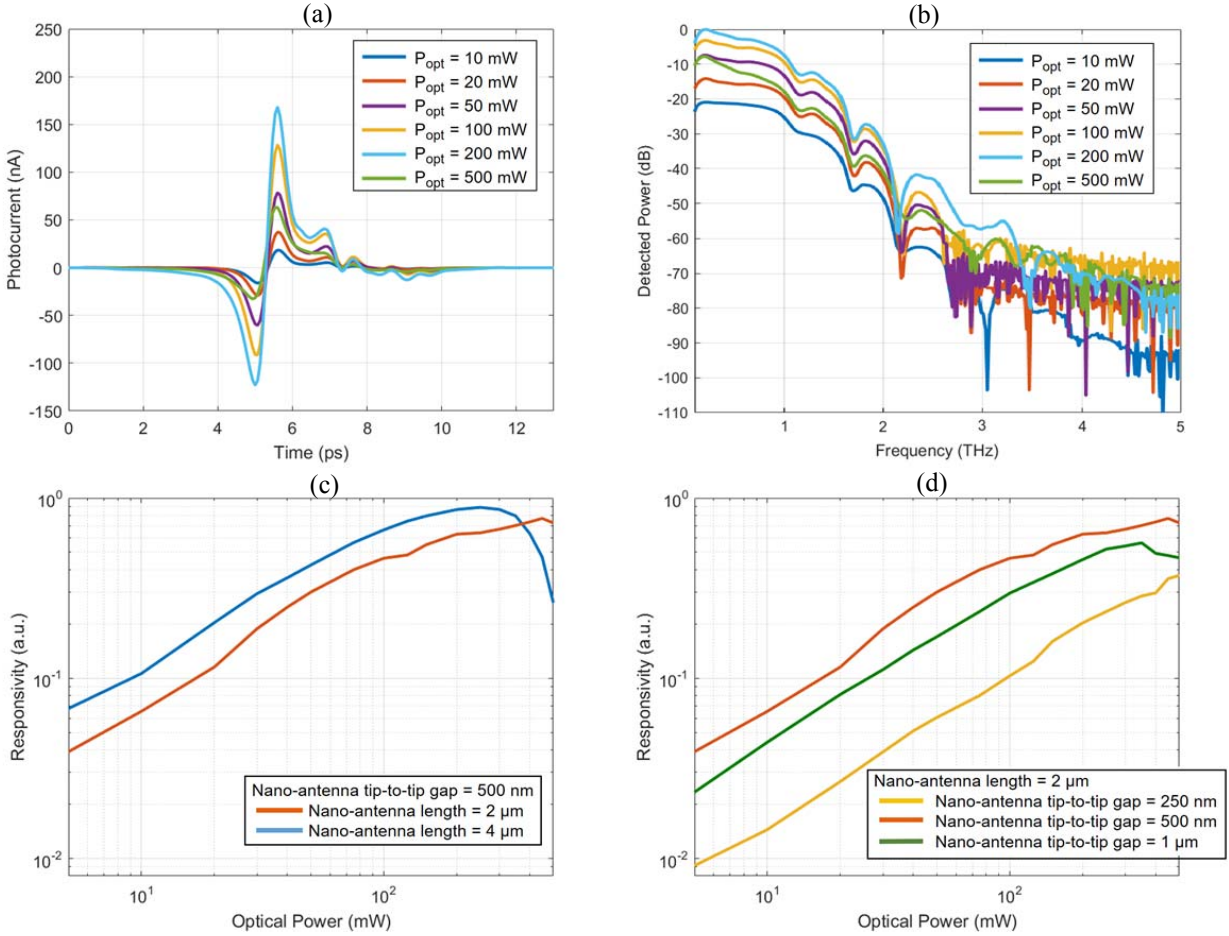

**Figure 5: Responsivity characteristics of the plasmonic nano-antenna arrays.** The measured photocurrent of a detector prototype with a nano-antenna length of 4 μm and tip-to-tip gap size of 500 nm in the time-domain and the corresponding power spectra are shown in (a) and (b), respectively. (c) Responsivity of detector prototypes with nano-antenna lengths of 2 μm and 4 μm and a tip-to-tip gap size of 500 nm at 1 THz, (d) responsivity of detector prototypes with a nano-antenna length of 2 μm and nano-antenna tip-to-tip gap sizes of 250 nm, 500 nm, and 1 μm at 1 THz.

The same time-domain terahertz spectroscopy system is used to measure noise characteristics of the detector prototypes. The optical beam pumping the terahertz emitter is blocked and the detector noise current is measured from the root mean square (rms) value of the induced photocurrent. The measurement results show that the detector noise current does not have a considerable dependence on the nano-antenna length (Fig. 6a), but has a strong dependence on the nano-antenna tip-to-tip gap size (Fig. 6b). This is because the primary noise source for the presented plasmonic photoconductive detector is the Johnson-Nyquist noise [29-31], with a noise current inversely proportional to the square root of the device resistance. Among the three detector prototypes with a nano-antenna length of 2 μm, the design with a nano-antenna tip-to-tip gap size of 500 nm offers the highest resistance and lowest noise current, consequently. Similarly, the design with a nano-antenna tip-to-tip gap size of 250 nm offers the lowest resistance and highest noise current, consequently. Since the device resistance is inversely proportional to the optical pump power level, a linear relation is observed between the noise current and the square root of the optical pump power.

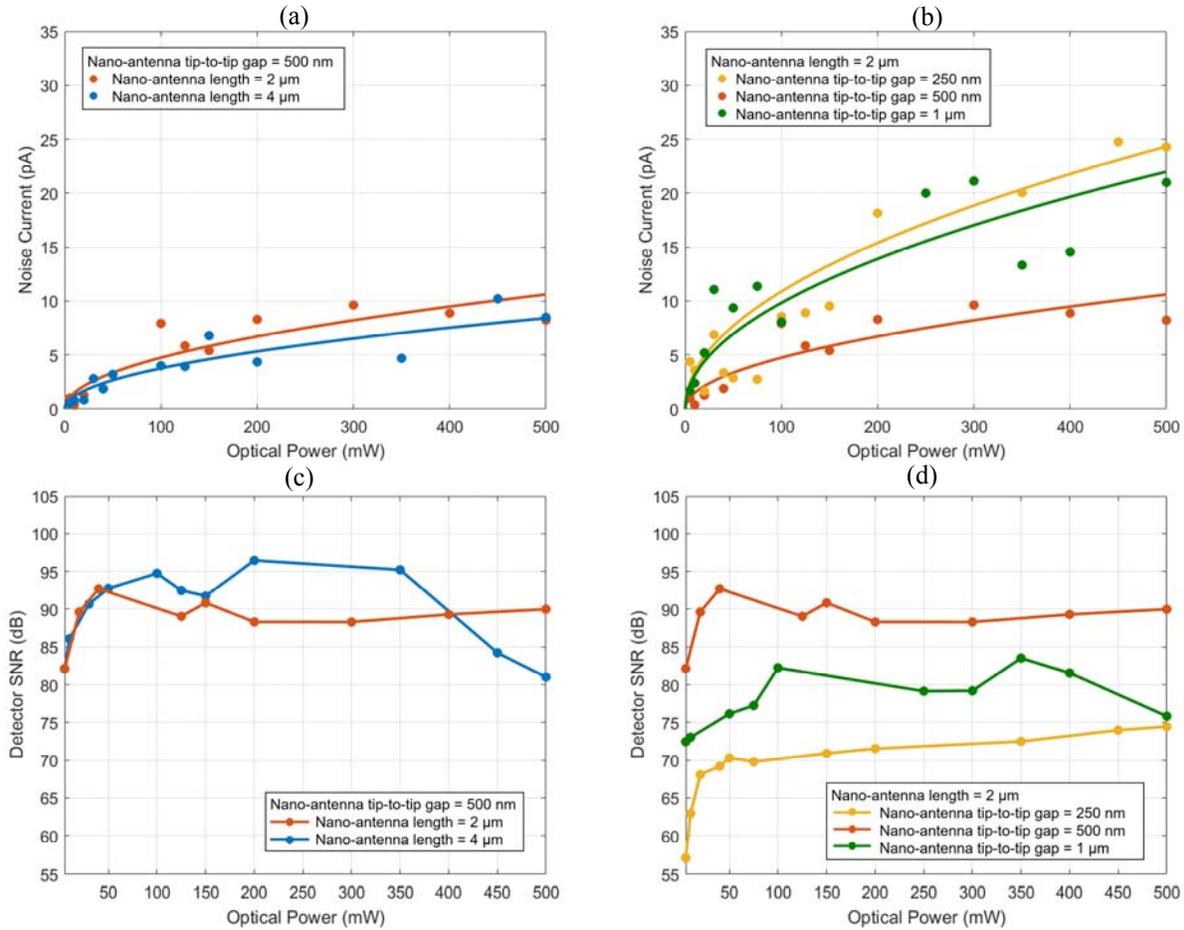

**Figure 6: Noise characteristics of the plasmonic nano-antenna arrays.** (a) The measured noise current of the detector prototypes with nano-antenna lengths of 2 μm and 4 μm and a tip-to-tip gap size of 500 nm, (b) the measured noise current of the detector prototypes with a nano-antenna length of 2 μm and nano-antenna tip-to-tip gap sizes of 250 nm, 500 nm, and 1 μm, (c) the measured SNR of the detector prototypes with nano-antenna lengths of 2 μm and 4 μm and a tip-to-tip gap size of 500 nm, (d) the measured SNR of the detector prototypes with a nano-antenna length of 2 μm and nano-antenna tip-to-tip gap sizes of 250 nm, 500 nm, and 1 μm. For each time-domain photocurrent data point used for calculating the detector noise current and SNR, 35 measurements are taken and averaged with an integration time of 10 ms.

Signal-to-noise ratio (SNR) of the detector prototypes is calculated from the measured photocurrent and noise current data. A relatively linear relation between the SNR and optical pump power is observed at optical pump power levels below 50 mW. This is because the detector responsivity is proportional to the optical pump power and the noise current is proportional to the square root of the optical pump power. A reduced increase in SNR values is observed at higher optical pump power levels due to the carrier screening effect. The designs with longer nano-antennas (Fig. 6c) and wider nano-antenna tip-to-tip gaps (Fig. 6d), which are more susceptible to the carrier screening effect, exhibit a roll-off in their SNR values at optical pump powers exceeding 200 mW. The highest SNR values are offered by the detector prototype with a nano-antenna length of 4 μm and tip-to-tip gap size of 500 nm due to its superior responsivity and noise performance. The SNR values can be further improved by increasing the number of the measured data points and data acquisition times [15]. SNR values as high as 107 dB are offered by the detector prototype with a nano-antenna length of 4 μm and a tip-to-tip gap size of 500 nm at 170 mW optical pump power when capturing and averaging 750 photocurrent data points with an integration time of 10 ms (Fig. 7). The SNR enhancement is mainly due to noise current reduction from 5 pA to 1.3 pA when increasing the data points from 35 to 750 [15].

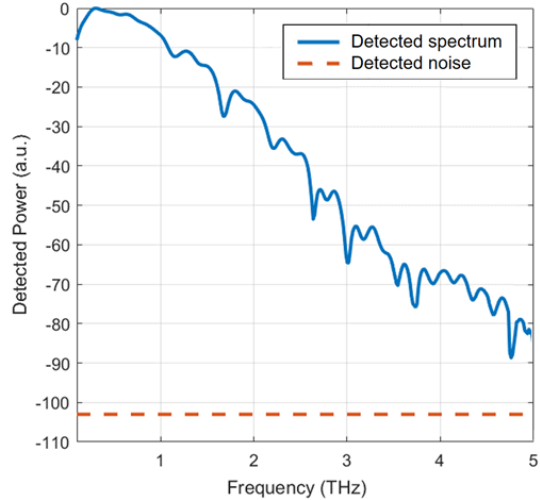

**Figure 7: Detection spectrum of the plasmonic nano-antenna arrays.** The measured power spectrum by the detector prototype with a nano-antenna length of 4 μm and a tip-to-tip gap size of 500 nm relative to the measured noise power at 170 mW optical pump power. The presented data are achieved by capturing and averaging 750 photocurrent data points with an integration time of 10 ms.

In summary, we present a high-performance terahertz detector based on plasmonic photoconductive nano-antenna arrays. It is specifically designed to allow a strong interaction between an incident terahertz and optical pump beam across a large device active area and over a broad range of terahertz frequencies. The large device active area allows robust operation without tight constraints on optical alignment and focusing. In order to find optimum designs, the impact of the detector geometry on its responsivity and noise characteristics is analyzed numerically and experimentally. We demonstrate broadband terahertz detection with SNR levels as high as 107 dB. Such high-performance terahertz detectors could extend scope and potential use of compact time-domain terahertz spectroscopy systems for many imaging and sensing applications.

**METHODS**
**Plasmonic nano-antenna array fabrication process.** The fabrication process starts with patterning the plasmonic nano-antennas through electron beam lithography, followed by Au deposition and liftoff. The Au lines connecting the nano-antennas are patterned next by optical lithography, followed by metal deposition and liftoff. The $Si_3N_4$ antireflection coating layer is added next by using plasma-enhanced chemical vapor deposition. The shadow metals are also patterned by optical lithography, followed by metal deposition and lift-off. Finally, contact vias are opened by reactive ion etching of the $Si_3N_4$ layer.

**ACKNOWLEDGEMENT**

The authors gratefully acknowledge the financial support from Moore Inventor Fellowship, Presidential Early Career Award for Scientists and Engineers (# N00014-14-1-0573), and NIH (#1R01GM112693-01A1).

**AUTHOR CONTRIBUTIONS STATEMENT**

N.T. Yardimci performed electromagnetic simulations, fabricated the device prototypes and performed device characterization. M. Jarrahi organized and supervised the project. Both authors discussed the results and commented on the manuscript.

**AUTHOR INFORMATION**

The authors declare no competing financial interests. Correspondence and request for materials should be addressed to Mona Jarrahi (mjarrahi@ucla.edu).